# Pseudocrystalline state of the surface of free xenon clusters

E.V. Gnatchenko, A.N. Nechay, A.A. Tkachenko, and V.N. Samovarov

*B. Verkin Institute for Low-Temperature Physics and Engineering*
*of the National Academy of Sciences of Ukraine*
*47, Lenin Ave., Kharkiv, 61103, Ukraine*
*E-mail: nechay@ilt.kharkov.ua*

**Abstract**

The paper presents the results of the experimental study of polarization bremsstrahlung profile halfwidth performed on free xenon clusters with various atom numbers. We used 0.7- and 0.3-keV electrons to excite mainly the core and the surface of the clusters, respectively. The halfwidth vs. atom number dependencies were found to be quite different for the 0.7- and 0.3-keV electrons. Analysis of the observed difference allowed us to conclude that the clusters with $N \approx$ 2000-8000 atoms per cluster ($R \approx$ 30-50 Å) consisted of a crystalline fcc core covered by a noncrystalline shell. Our data provide evidence that the surface shell is in the *pseudocrystalline state*, which is its first observation in rare gas clusters. The pseudocrystalline state observed in metallic clusters is a volume effect, the clusters in this state are structurally unstable and fluctuate continuously between liquid and solid phases.

**1. Introduction**

Clusters are systems with a finite number of constituent particles characterized by a large fraction of surface particles. As a result, the surface subsystem has a great effect on the physicochemical properties of clusters. The structure of the cluster surface may differ considerably from that of the inner layers. A number of papers have been dedicated to the investigation of the surface melting of clusters implying the possibility for the surface layers of Lennard-Jones and metallic clusters to be in a quasi-liquid (liquid) state at temperatures below their melting temperature (*premelting effect*) [1-4]. Experiments on surface melting are carried out mainly with metallic clusters. Of special interest are papers dealing with the so called *pseudocrystalline* (PC) state in such clusters [4, 5-8]. Transition to the PC state can be listed among the cluster effects, since it only takes place in clusters whose radius is below some critical value $R_c$ [7-8]. The surface layers of clusters can be in the PC state too [8]. A system in this state is structurally unstable and fluctuates continuously between liquid and solid phases.

The surface state of rare gas clusters has not been studied experimentally yet. The purpose of this paper is to get more information about the structure of the surface layers of substrate-free xenon clusters produced by the method of adiabatic expansion of gas flowing through a supersonic nozzle into a vacuum.

Polarization bremsstrahlung (PBS) spectroscopy has been used here for the first time to study the structural state of the cluster surface and core. Polarization bremsstrahlung is generated by the oscillations of the dipole moment of an atom induced by the alternating field of an incident electron (or other charged particle) interacting with a gas or solid target [9]. In contrast to the featureless spectrum of ordinary bremsstrahlung, the PBS spectrum has a pronounced resonance in the ultrasoft x-ray spectral region 60-200 eV ($\lambda$ = 207-62 Å) [9-11].

Our previous experiments [12] showed that when xenon clusters were excited by electrons with the energy of 0.7 keV, the PBS profile resulted substantially narrower for crystalline clusters than for icosahedral ones, which suggests that PBS profile can be used to study the structural state of clusters.

In the present work, we used not only the *high-energy* 0.7-keV electrons, but also *low-energy* electrons with the energy 0.3 keV, whose contribution to the excitation of the cluster surface is considearbly greater. We have found that the dependencies $W(N)$ of PBS halfwidths on number of atoms in the cluster differ greatly for the 0.3- and 0.7-keV electrons. Analysis of the differences allows us to conclude that the crystalline fcc core of the clusters is covered by a noncrystalline shell in the cluster size range $\approx$ 2000-8000 atoms/cluster ($R \approx$ 30-50 Å).

The nature of the noncrystalline shell is considered from the viewpoint of the two possible surface structures: (i) *glassy state* and (ii) *pseudocrystalline state*. It is shown that the cooling velocities ($\approx 2 \cdot 10^7$ K/s) of liquid droplets and "hot" solid clusters used in our experiments are not sufficient to create a glassy state on the cluster surface. It is suggested for the first time for rare gas clusters that the surface shell of such clusters is in the PC state.

## 2. Experiment

The experimental method for studying PBS with intermediate-energy electrons scattered by atomic and cluster beams was described in detail in Refs. [11-12]. Below, we are mentioning the main particularities of the present experiments. The spectra were registered in the energy range 80-180 eV by a spectrometer-monochromator and a proportional counter with an error below 3 eV. Atomic and cluster beams of xenon were produced by the method of adiabatic expansion of gas flowing into a vacuum through a supersonic nozzle with the following parameters: total opening angle of the cone $2\alpha = 9.5°$, critical cross-section diameter 0.3 mm, ratio of the exit cross-section to the critical cross-section $S_{ex} = 59.3$. By decreasing temperature of the gas at the nozzle entrance $T_0$ (450-180 K) and increasing its pressure $P_0$ (0.3 -1 atm) it was possible to transform atomic beams to cluster ones and change the cluster size. Our experiments were performed with the three types of beams: atomic beams, beams containing icosahedral clusters, and those containing clusters (up to $N \approx 12000$ atoms per cluster, $R \approx 56$ Å) with a crystalline fcc core. The average number of atoms in the cluster $N$ was calculated with an accuracy of 30-50% using the Hagena relations $N = \varphi(P_0, T_0)$, which also take into account the geometric parameters of the nozzle (see, for example, Ref. [13]). The temperature of the xenon clusters in the supersonic beams, estimated from the measured electron-diffraction data of lattice parameter, was $T_{cl} = 79 \pm 8$ K [14, 15].

Analysis of the energy release in solid xenon performed in this work has shown that electrons with the energy 0.3 keV are more efficient in exciting the cluster surface than 0.7-keV electrons.

## 3. Results

Figure 1 shows the PBS spectra in the coordinates $\omega d\sigma/d\omega$ ($d\sigma/d\omega$ being PBS differential cross-section) versus photon energy $\hbar\omega$ for clusters with $N \approx 8000$ atoms/cluster excited by 0.3- and 0.7-keV electrons. The smooth lines show the averaged experimental data. It can be seen that the PBS maximum for 0.7-keV electrons is noticeably narrower than that for 0.3-keV electrons.

Figure 2 shows the $W(N)$ dependencies of PBS profile halfwidth on atom number in the cluster for electrons with the energies 0.3 and 0.7 keV. The dependencies are normalized to the PBS profile halfwidth for atomic beam. It can be seen that they differ from each other: in the case of 0.7-keV electrons, the halfwidth narrows abruptly for clusters with over 1000-1500 atoms/cluster by up to $\approx 30\%$ for $N \approx 12000$ atoms/cluster. The inset to Fig. 2 shows the behavior of $W(N)$ for 0.7-keV electrons and $N < 1000$ atoms/cluster when the profile halfwidth changes insignificantly. Clusters of such dimensions are noncrystalline and have an icosahedral structure [15, 16]. Crystalline clusters of xenon with an fcc core are formed in the beam if the atom number in the cluster exceeds $\approx 1500$ atoms/cluster ($R \approx 27$ Å) [16]. We should note that, according to the electron diffraction data, fcc clusters of rare gases are monodomains at $R < 100$ Å [17]. Therefore, the strong narrowing of PBS profile observed for 0.7-keV electrons occurs when they interact mainly with the crystalline fcc core of the clusters. At the same time, in the case of 0.3-keV electrons, the PBS profile halfwidth is virtually the same for atomic beams, noncrystalline and crystalline (with up to 8000 atoms) clusters. This allows us to conclude that the fcc xenon clusters with $N \approx 2000 - 8000$ atoms/cluster have a noncrystalline surface shell too.

## 4. Discussion

*Glassy state.* Formation of the glassy state depends much on the rate at which a system of particles is cooled.

For the glassy state to be formed in the cluster surface, the following quite evident inequality must be satisfied: $\tau < \tau_D$, where $\tau_D$ is a characteristic time of atom motion in the quasiliquid surface layer near the xenon triple point $T_{tr} = 161$ K, while $\tau$ is a characteristic time of cooling. We have estimated the times and showed that for the cooling rate of $2 \cdot 10^7$ K/s, achieved in our experiments, the characteristic time of cooling $\tau$ is about $8 \cdot 10^{-6}$ s. The characteristic diffusion time near the triple point $T_{tr}$ is $\tau_D \approx 10^{-9}$ s for a surface layer of thickness $\ell = 10$ Å, which means that a very strong inequality $\tau_D \ll \tau$ is satisfied. No glassy state was therefore formed in the surface shell of the clusters studied in our experiments.

*Pseudocrystalline state*. Let us discuss the other possible state of the cluster surface layer. It is known that as the cluster size decreases, the cluster melting temperature also decreases. For clusters with radius $R$, this size effect can be expressed by the following relation (in the approximation of equal densities in the solid and liquid phases, $\rho_s \approx \rho_\ell$) [18]:

$$\frac{T_m^{cl}}{T_m} = \left(1 - \frac{2(\sigma_s - \sigma_l)}{q\rho_s R}\right) = \left(1 - \frac{R_0}{R}\right), \tag{1}$$

$T_m$ and $T_m^{cl}$ being the melting temperatures of solid crystal and cluster, respectively. In the case of xenon, the values in Eq. (1) are as follows: surface tension of an fcc crystal is $\sigma_s = 61$ erg/cm$^2$ (80 K) [19]; surface tension of the liquid phase is $\sigma_\ell = 18.8$ erg/cm$^2$ [20]; specific heat of fusion is $q = 3.7 \cdot 10^{-14}$ erg [20]; density in the solid phase is $\rho_s = 1.7 \cdot 10^{22}$ cm$^{-3}$ (80 K) [20]. It follows that the parameter $R_0$ in Eq. (1) is 13 Å and the melting temperature is $T_m^{cl} = 101$ K for clusters with 3000 atoms/cluster ($R \approx 35$ Å) and $T_m^{cl} = 113$ K for those with 6000 atoms/cluster ($R \approx 44$ Å).

We should note that stacking faults typical of substrate-free fcc clusters lower cluster melting temperature. In xenon clusters with $N = 3000$ and 6000 atoms/cluster, their density, according to the electron diffraction data, is rather high, the ratio of the number fault planes to the total number of planes being 0.06 and 0.05, respectively, every cluster containing on the average four fault planes intersecting each other [16]. Under these conditions, the cluster melting temperature may be quite close to their temperature 80 K and one can therefore talk in terms of surface melting of clusters.

Surface melting of bulk crystals has long been studied both theoretically and by various experimental techniques in metals, molecular media, and solid rare gases (see, for example, Ref. [21]). It was shown experimentally and theoretically in Ref. [22] that the surface faces (111)

6and (100) in solid xenon start melting at $T_s = 0.8T_m$ (129 K), where $T_m$ is the temperature of homogeneous melting of xenon being equal to 161.4 K. Starting from $T_s$, deeper faces become unstable and an interface is formed between the crystalline phase and the liquid (quasiliquid) layer. Surface melting, as has already been said, is widely studied in metallic clusters. It has been found, for example, that as substrate-free Sn clusters with the radii 50-100 Å are heated up, the liquid surface layer grows to occupy a rather great portion of the cluster volume, its thickness being 18 Å [2].

We would like to note that the surface layer is liquid only in those clusters which have a radius exceeding some critical value $R_c$ [7, 8]. If the cluster radius is below $R_c$, the surface layer can enter the pseudocrystalline state [8].

The experimental studies undertaken to find the PC state in metallic clusters have laid the foundation of a new physics of small objects. One of the first papers aimed in this direction dealt with electron microscopy studies of gold clusters of about 20 Å in radius [5] (the theoretical value being $R_c = 37$ Å [8]). It was noticed in the paper that within the traditional concept of matter the observed state could be attributed neither to liquid nor to solid phase. In the cluster phase diagram constructed for different cluster sizes, the region of PC state lies between the regions of liquid and crystalline states, as well as of a noncrystalline state with an icosahedral structure [6]. The system fluctuates continuously between these different states, since the large entropy of small clusters prevents from stabilization of only one state [23]. As the cluster radius and temperature grow, the cluster size range characterized by the PC state shrinks rapidly [6].

The existence of a critical cluster size is due to the following reason. As it has already been mentioned, the cluster melting temperature $T_m^{cl}$ decreases with decreasing cluster size. The temperature of surface melting $T_s$ is a weaker function of cluster radius. Consequently, at $R < R_c$, the inequality $T_s > T_m^{cl}$, defining the interval of existence of a PC layer, will hold true. For example, for xenon clusters with 6000 atoms/cluster (see above), the melting temperature is $T_m^{cl}$ = 113 K, but $T_s$, under the assumption that in the cluster it is almost the same as in a bulk sample, turns out to be 129 K, which is higher than $T_m^{cl}$.

Using the theoretical paper [8] to calculate the critical radius for xenon clusters, we can find the value $R_c$ = 55 Å. For comparison purposes, we would like to note that for fusible metals the calculated values based on the theoretical results of Ref. [8] are as follows: $R_c$ = 52 Å (Pb clusters), $R_c$ = 94 Å (Sb clusters), $R_c$ = 38 Å (Sn clusters; the experimental values being 25 Å [7] and 35 Å [24]). Therefore, in xenon fcc clusters with 2000-8000 atoms/cluster ($R \approx$ 30-50 Å) studied in this paper, there may appear a PC surface layer persisting up to the complete melting of the clusters. This layer can account for the weak dependence of PBS profile halfwidth on atom number observed in such clusters for 0.3-keV electrons. But for $N$ > 8000 atoms/cluster (R > 50 Å), the PBS profile (0.3-keV electrons) starts narrowing sharply. In the cluster size range 8000-12000 atoms/cluster, its halfwidth decreases by almost 25% with cluster radius growing by only 10%. Xenon clusters in this cluster size range have the radii that exceed the critical value and one can thus believe that the structural state of their surface layers changes.

## 5. Conclusions

We studied experimentally polarization bremsstrahlung profile halfwidth $W(N)$ as function of atom number in substrate-free xenon clusters excited by electrons with the energies 0.7 and 0.3 keV used to probe into the cluster core and surface, respectively. It was found that the $W(N)$ dependencies behave differently in the case of 0.7- and 0.3-keV electrons. We have analyzed the difference and concluded that the clusters with $N \approx$ 2000-8000 atoms/cluster ($R \approx$ 30-50 Å) have a crystalline fcc core covered by a noncrystalline surface layer. We have considered the nature of this layer from the viewpoint of the possible formation of a glassy and pseudocrystalline states. The analysis of our supports the existence of the surface pseudocrystalline state, which has never been observed in rare gas clusters before.

The authors are grateful to Dr. V.L. Vakula for the fruitful discussions and assistance in preparing the paper.

*The full-length version of the paper will be published in Low Temperature Physics in December, 2012.*

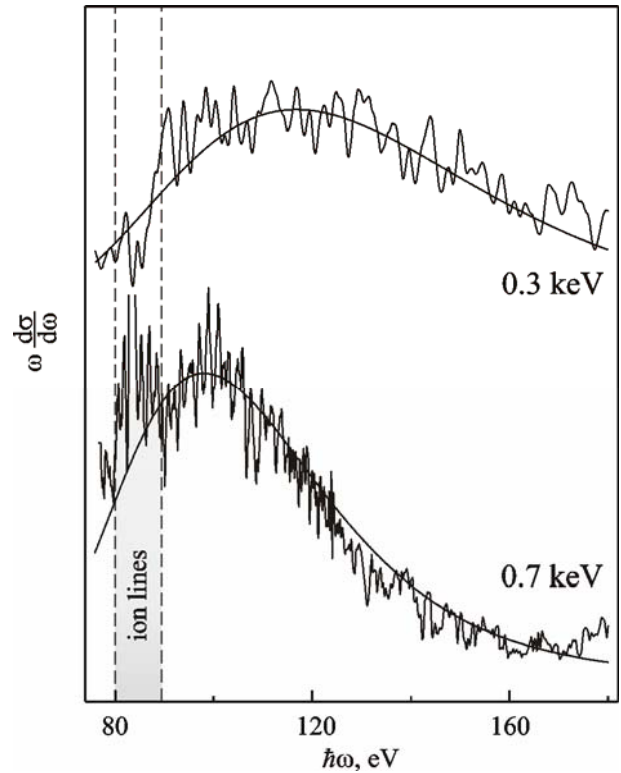

Fig. 1. Polarization bremsstrahlung spectra from a beam of xenon clusters with an average cluster size $N \approx 8000$ atoms/cluster excited by 0.3- and 0.7-keV electrons. The result of the data averaging is shown by smooth curves. The shaded area shows the spectral region where emission lines of xenon ions overlap with the PBS spectrum.

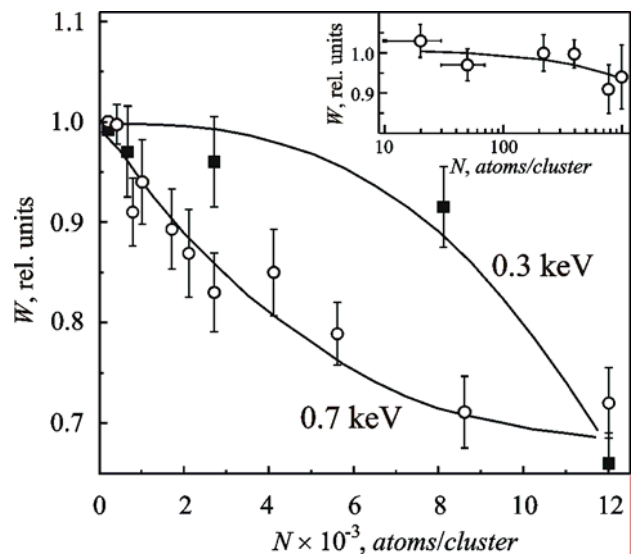

Fig. 2. Reduced PBS profile halfwidths as function of number of atoms in the cluster for electrons with the energies 0.3 (■) and 0.7 keV (o). The inset shows the $W(N)$ dependence for 0.7-keV electrons and $N < 1000$ atoms/cluster.